\documentclass[11pt,a4paper]{article}
\usepackage{jheppub}
\usepackage{amsmath}
\usepackage{graphicx}
\usepackage{slashed}

\def\coeff#1#2{{\textstyle {\frac {#1}{#2}}}}

\def\L{\pounds}

\def\E{{\rm \scriptscriptstyle E}}

\def\k{{\bf k}}
\def\x{{\bf x}}

\title{Kubo formulas for thermodynamic transport coefficients}
\author{Pavel Kovtun}
\author{and Ashish Shukla}
\affiliation{Department of Physics \& Astronomy,  University of Victoria,
PO Box 1700 STN CSC, Victoria, BC,  V8W 2Y2, Canada}

\emailAdd{pkovtun@uvic.ca}
\emailAdd{ashish@uvic.ca}

\abstract{
Uncharged relativistic fluids in 3+1 dimensions have three independent thermodynamic transport coefficients at second order in the derivative expansion. Fluids with a single global $U(1)$ current have nine, out of which seven are parity preserving. We derive the Kubo formulas for all nine thermodynamic transport coefficients in terms of equilibrium correlation functions of the energy-momentum tensor and the current. All parity-preserving coefficients can be expressed in terms of two-point functions in flat space without external sources, while the parity-violating coefficients require three-point functions. We use the Kubo formulas to compute the thermodynamic coefficients in several examples of free field theories.
}

\begin{document}
\maketitle

\section{Introduction}
When a macroscopic system in equilibrium is subject to external fields, it reacts to the fields by adjusting its pressure, energy density, and other thermodynamic functions. If the system is subject to external electric and magnetic fields, the response is described by the electric and magnetic susceptibilities, which determine the electric permittivity and the magnetic permeability of matter. If the system is subject to external gravitational fields, the analogous gravitational susceptibilities determine the appropriate response of the free energy. 

When the same system is perturbed out of equilibrium, the equilibrium susceptibilities can contribute to non-equilibrium phenomena. For example, the electric and magnetic susceptibilities determine the speed of light in matter, while the pressure determines the speed of sound in matter, through the equation of state.

Motivated by the applications of relativistic hydrodynamics to the hot sub-nuclear matter~\cite{Gale:2013da, Jeon:2015dfa}, we will be focusing on relativistic fluids in this paper. The most basic thermodynamic susceptibility is of course the pressure itself, which can be viewed as a response of the free energy to a diagonal metric perturbation thanks to the covariant definition of the energy-momentum tensor in relativistic systems.

For relativistic fluids in 3+1 dimensions in curved space, the transport coefficients that are thermodynamic in nature were first noticed in \cite{Baier:2007ix, Bhattacharyya:2008jc} at second order in the derivative expansion, though their thermodynamic nature was not fully appreciated at the time. For fluids in 2+1 dimensions, analogous thermodynamic transport coefficients already appear at first order in the derivative expansion~\cite{Jensen:2011xb}. They have been variously referred to in the literature as ``thermodynamic response parameters'', ``thermodynamic transport coefficients'', ``thermodynamical hydrodynamic coefficients'', ``equilibrium hydrodynamic coefficients'', or ``non-dissipative transport coefficients''.%
\footnote{
One should keep in mind that not all non-dissipative transport coefficients are thermodynamic in nature. As an example, Hall viscosity is a non-dissipative, non-thermodynamic transport coefficient.
}
There is a multitude of notations for these coefficients in the literature, and the translation between different conventions is not always straightforward.

While the thermodynamic transport coefficients were first noticed in the context of hydrodynamics~\cite{Baier:2007ix, Bhattacharyya:2008jc, Jensen:2011xb}, their connection with thermodynamics was not explored until \cite{Banerjee:2012iz, Jensen:2012jh}. These papers showed that the relevant coefficients in the constitutive relations follow from the equilibrium partition function, including the highly non-trivial constraints~\cite{Jensen:2011xb, Bhattacharyya:2012nq} demanded by the local positivity of entropy production. We will refer to the thermodynamic coefficients that appear in the constitutive relations as ``thermodynamic transport coefficients'', and to the coefficients in the equilibrium free energy as ``thermodynamic susceptibilities''. Thermodynamic transport coefficients are linear combinations of thermodynamic susceptibilities and their derivatives~\cite{Banerjee:2012iz, Jensen:2012jh}. In the classification of non-dissipative transport coefficients in~\cite{Haehl:2015pja}, thermodynamic transport coefficients correspond to class H\textsubscript{S}.%
\footnote{
Class H\textsubscript{S} is a subclass within class L, with L = H\textsubscript{S} $\cup$ $\bar{\text{H}}_{\text{S}}$. Class L comprises non-dissipative transport which admits description in terms of a local Lagrangian. The formalism we discuss below does not immediately translate to transport coefficients in the class $\bar{\text{H}}_{\text{S}}$.}

We will be considering fluids with a conserved global $U(1)$ charge, such as the baryon number. We will refer to the fluids that can be locally described as having a temperature and a chemical potential for the global $U(1)$ charge as ``charged fluids''. The system can be coupled to the corresponding non-dynamical external $U(1)$ gauge field, and to the non-dynamical external metric. The thermodynamic susceptibilities then include the usual ``electric'' and ``magnetic'' susceptibilities, as well as the response of the free energy to the vorticity, to the Riemann curvature, the magneto-vortical response, etc. In 3+1 dimensions, there are nine such susceptibilities at two-derivative order~\cite{Banerjee:2012iz}. These susceptibilities will appear in the constitutive relations and in equilibrium correlation functions for fluids in flat space without external $U(1)$ fields. 

As we will see later, the second-order thermodynamic transport coefficients in QCD at non-zero baryon number chemical potential (in flat space without external fields) are determined by five thermodynamic susceptibilities. For a parity-preserving conformal theory, the second-order thermodynamic transport coefficients at non-zero chemical potential (in flat space without external fields) are determined by three thermodynamic susceptibilities. 

Kubo formulas for second-order thermodynamic transport coefficients were derived in \cite{Moore:2010bu, Arnold:2011ja, Moore:2012tc} for uncharged fluids. Further, \cite{Moore:2012tc, Romatschke:2009ng} evaluated these coefficients for non-interacting massless scalars, fermions, and gauge fields. Ref.~\cite{Megias:2014mba} evaluated the seven parity-even thermodynamic susceptibilities of a charged fluid in a theory of free massless fermions, using a dimensionally reduced partition function in curved space. In addition to the above functional methods, \cite{Buzzegoli:2017cqy} evaluated the thermodynamic second-order coefficients for charged fluids of free scalars and fermions using operator methods.

Our focus in this paper will be on the Kubo formulas for all nine second-order susceptibilities in 3+1 dimensions. 
We will write down the Kubo formulas for the susceptibilities, rather than for the thermodynamic transport coefficients. This is natural, as the thermodynamic transport coefficients are derived objects, while the susceptibilities are fundamental. We will find that the Kubo formulas for all seven parity-preserving susceptibilities can be written in terms of two-point correlation functions of the energy-momentum tensor and the $U(1)$ current. In other words, using three-point functions to evaluate the thermodynamic transport coefficients as in~\cite{Moore:2012tc, Buzzegoli:2017cqy} is not necessary, and two-point functions are sufficient.%
\footnote{
  Appendix D of \cite{Chapman:2013qpa} mentions this point for uncharged fluids. Also, we are not aware of a systematic method to predict the minimum number of operator insertions needed to compute a given transport coefficient; it appears that one has to do this analysis independently at each order in the derivative expansion.
}
Using two-point functions will hopefully allow for an easier evaluation of these transport coefficients on the lattice~\cite{Moore:2012tc, Philipsen:2013nea}, and in holography~\cite{Baier:2007ix, Bhattacharyya:2008jc, Finazzo:2014cna}.
Further, we write the free energy in a covariant form as in \cite{Jensen:2012jh}, which directly gives covariant expressions for the energy-momentum tensor and the $U(1)$ current, generalizing the results of \cite{Jensen:2012jh} for second-order transport coefficients to charged fluids. (See \cite{Bhattacharyya:2014bha} for the generalization of the results of \cite{Banerjee:2012iz} to charged fluids). We illustrate our Kubo formulas by evaluating the susceptibilities in a few examples of free field theories. 

The paper is organized as follows. In section~\ref{SS:EGF} we introduce the equilibrium generating functional for thermodynamic correlation functions, and define the energy-momentum tensor as well as the conserved current in terms of its variation under external sources. In section~\ref{SS:DE} we write the generating functional in terms of the nine independent susceptibilities appearing at second order in the derivative expansion. Section~\ref{SS:TA} briefly talks about the trace anomaly. Section~\ref{SS:EMTC} then provides the expressions for the energy-momentum tensor and the conserved current in terms of the second order susceptibilities for a charged fluid in the absence of external fields. We present the Kubo formulas in section~\ref{SS:KF}, and express the susceptibilities in terms of equilibrium two- and three-point functions. In sections~\ref{SS:Scalars}, \ref{SS:Fermions} and \ref{SS:Gauge} we evaluate the susceptibilities for free scalar, free Dirac fermion, and free gauge fields. We end with a discussion in section~\ref{SS:Discussion}. Appendix \ref{SS:TOC} provides relations between our thermodynamic susceptibilities and the ones that have appeared previously in literature.

\section{Thermodynamics in external fields}
\label{sec:thermodynamics}
\subsection{Equilibrium generating functional}
\label{SS:EGF}
We consider a macroscopic system that has degrees of freedom which couple to the external metric $g_{\mu\nu}$ and to an external Abelian gauge field $A_\mu$.
The matter in equilibrium is described by the generating functional for equilibrium (zero-frequency) correlation functions~$W[g,A]$. Such generating functionals have been discussed starting from~\cite{Banerjee:2012iz, Jensen:2012jh}, with applications to relativistic hydrodynamics. We follow the presentation of \cite{Jensen:2012jh, Jensen:2013kka, Kovtun:2016lfw}. Equilibrium is characterized by a timelike Killing vector which we denote by $V$. The coordinates in which $V^\mu = (1,{\bf 0})$ correspond to the matter at rest.\footnote{ We use the mostly-plus convention for the metric.} The matter velocity, temperature, and the chemical potential are defined as
\begin{equation}
\label{eq:uTmu}
  u^\mu = \frac{V^\mu}{\sqrt{-V^2}}\,,\ \ \ \ 
  T = \frac{1}{\beta_0 \sqrt{-V^2}}\,,\ \ \ \ 
  \mu = \frac{ V^\mu A_\mu + \Lambda_V}{\sqrt{-V^2}}\,,
\end{equation}
where $\beta_0$ is a constant which sets the normalization of temperature, and $\Lambda_V$ is a gauge function which ensures that the chemical potential is gauge invariant.
Denoting the Lie derivative by $\L$, the conditions for being in equilibrium are
\begin{equation}
\label{eq:equil-cond}
  \L_V g_{\mu\nu}=0\,,\ \ \ \ \L_V A_\mu + \partial_\mu \Lambda_V = 0\,.
\end{equation}

For systems with a finite correlation length, the equilibrium generating functional is extensive in the thermodynamic limit, and can be written as
\begin{equation}
\label{eq:W1}
  W[g,A] = \int\!\!d^{d+1}x\,\sqrt{-g}\; {\cal F}[g,A]\,,
\end{equation}
where the density ${\cal F}[g,A]$ is a local function of the external sources $g_{\mu\nu}$ and $A_\mu$. We will assume that the microscopic theory has no chiral anomalies, and so $W[g,A]$ is gauge- and diffeomorphism-invariant. We define the energy-momentum tensor $T^{\mu\nu}$ and the current $J^\mu$ in the standard fashion,
\begin{equation}
\label{eq:TJ-def}
  \delta W[g,A] = \coeff12 \int\!\!d^{d+1}x\,\sqrt{-g}\, T^{\mu\nu} \delta g_{\mu\nu} + 
  \int\!\!d^{d+1}x\, \sqrt{-g} \, J^\mu \delta A_\mu\,.
\end{equation}
The diffeomorphism- and gauge-invariance of $W[g,A]$ imply, respectively,
\begin{subequations}
\label{eq:TJ-cons-1}
\begin{align}
  & \nabla_{\!\mu} T^{\mu\nu} = F^{\nu\lambda}J_\lambda\,,\\
\label{eq:J-cons-1}
  & \nabla_{\!\mu} J^\mu = 0\,,
\end{align}
\end{subequations}
where the $U(1)$ gauge field strength is $F_{\mu\nu}=\partial_\mu A_\nu - \partial_\nu A_\mu$.
When the external sources $g$ and $A$ vary in space on length scales much longer than the correlation length, the density ${\cal F}[g,A]$ can be written as a derivative expansion of the external sources. The problem of finding the generating functional then boils down to finding the gauge- and diffeomorphism-invariants made out of the metric, the gauge field, and the quantities in (\ref{eq:uTmu}), up to a given order in derivatives.

Before we start writing down the invariants that appear in (\ref{eq:W1}), it is worth emphasizing the identities which follow from the fact that the system is in equilibrium. The equilibrium conditions (\ref{eq:equil-cond})  imply that the fluid velocity, temperature, and the chemical potential defined by (\ref{eq:uTmu}) are not arbitrary, but rather must obey
\begin{subequations}
\label{eq:econstr}
\begin{align}
  & u^\lambda \partial_\lambda T = 0\,,\ \ \ \ u^\lambda \partial_\lambda \mu = 0\,,\\[5pt]
  & a^\lambda = -\Delta^{\lambda\nu}\partial_\nu T/T\,,\\[5pt]
\label{eq:ETmu}
  & E^\lambda = T\,\Delta^{\lambda\nu} \partial_\nu\!\left(\frac{\mu}{T}\right)\,,\\[5pt]
  & \nabla{\cdot}u = 0\,,\ \ \ \ \sigma^{\mu\nu} = 0\,.
\end{align}
\end{subequations}
Here the acceleration is $a^\mu \equiv u^\lambda\nabla_\lambda u^\mu$, the projector $\Delta^{\mu\nu} \equiv g^{\mu\nu} + u^\mu u^\nu$ projects onto the space orthogonal to $u^\mu$, the shear tensor is $\sigma^{\mu\nu} \equiv \Delta^{\mu\alpha} \Delta^{\nu\beta}(\nabla_\alpha u_\beta + \nabla_\beta u_\alpha -\coeff23 \Delta_{\alpha\beta} \nabla{\cdot}u)$, and the electric field is $E_\mu \equiv F_{\mu\nu} u^\nu$. The first equation in (\ref{eq:econstr}) says that $T$ and $\mu$ are time-independent in equilibrium. The second equation in (\ref{eq:econstr}) says that the gravitational potential induces a temperature gradient. This is a consequence of Tolman's law~\cite{Tolman:1930ona} (equilibrium temperature is proportional to $1/\sqrt{-g_{00}}$ in the appropriate coordinates). The third equation in (\ref{eq:econstr}) says that the electric field induces a charge gradient. This is a formal way to express the phenomenon of electric screening. Alternatively, if (\ref{eq:ETmu}) were not true, there would be entropy production due to the electrical conductivity. The last equation in (\ref{eq:econstr}) says that the expansion and shear must vanish in equilibrium. If it were not so, there would be entropy production due to the bulk and shear viscosities.

We will find it convenient in what follows to use the electromagnetic polarization tensor. As the density ${\cal F}[g,A]$ is local and gauge-invariant, one can formally consider it to be a function of $A_\mu$ and the field strength $F_{\mu\nu}$. We then have
$$
  \delta_{A,F}W = \int\!\! d^{d+1}x\,\sqrt{-g} \left[ J^\mu_{\rm f}\delta A_\mu 
  +\coeff12 M^{\mu\nu} \delta F_{\mu\nu}\right]\,,
$$
which defines the current $J^\mu_{\rm f}$ and the anti-symmetric polarization tensor $M^{\mu\nu}$. Of course, the exact way how one chooses to consider $W[A]$ as a function of $A_\mu$ and $F_{\mu\nu}$ is ambiguous. This ambiguity is the ambiguity of separating the charge/current into the components corresponding to ``bound charges'' and ``free charges''. While $J^\mu_{\rm f}$ and $M^{\mu\nu}$ are ambiguous, the total current $J^\mu$ defined by (\ref{eq:TJ-def}) is not, and is given by 
$$
  J^\mu = J^\mu_{\rm f} - \nabla_{\!\lambda} M^{\lambda\mu}\,.
$$
The first term can be called the current of free charges, and the second term the current of bound charges. A convenient choice of fixing the ambiguity in the definition of $J^\mu_{\rm f}$ is to use (\ref{eq:ETmu}) to trade the derivatives of the chemical potential in the density ${\cal F}[g,A]$ for the electric field. This gives $J^\mu_{\rm f} = \rho u^\mu$, where $\rho\equiv \partial{\cal F}/\partial\mu$ defines the density of free charges. Then
\begin{equation}
\label{eq:JM}
  J^\mu  = \rho u^\mu  - \nabla_{\!\lambda} M^{\lambda\mu}\,,
\end{equation}
to all orders in the derivative expansion. Note that $M^{\mu\nu}=-M^{\nu\mu}$, and the bound current does not contribute to the conservation equation (\ref{eq:J-cons-1}). See \cite{Kovtun:2016lfw} for more details about the electric and magnetic contributions to~$M^{\mu\nu}$.

\subsection{Derivative expansion}
\label{SS:DE}
We next specify the derivative counting. We choose the counting scheme in which the metric is $g_{\mu\nu}\sim O(1)$, so that the Riemann tensor is $O(\partial^2)$. Similarly, the temperature is $T\sim O(1)$. If the matter in question has degrees of freedom that carry ``electric'' charges (as would be in a  conductor), the chemical potential is also $\mu\sim O(1)$. The equilibrium condition (\ref{eq:ETmu}) then requires that the electric field is $E_\mu\sim O(\partial)$. In an insulator, on the other hand, $\mu$ is not a relevant thermodynamic variable, and one can take $E_\mu\sim O(1)$. We will be considering conducting matter without macroscopic $O(1)$ magnetic fields, and will take $A_\mu\sim O(1)$.

At zeroth order in derivatives we then have only two invariants, $T$ and $\mu$. Thus the generating functional is
$$
  W[g,A] = \int\!\!d^{d+1}x\,\sqrt{-g}\, p(T,\mu) +\dots\,,
$$
where $p(T,\mu)$ is some function of $T$ and $\mu$ (which is in fact the pressure), and the dots denote the terms of order $O(\partial)$ and higher. The functional form of $p(T,\mu)$ is to be determined from the microscopic theory. The energy-momentum tensor and the current which follow from the definitions (\ref{eq:TJ-def}) are
\begin{subequations}
\label{eq:TJ0}
\begin{align}
  & T^{\mu\nu} = \epsilon u^\mu u^\nu + p \Delta^{\mu\nu} + \dots\,,\\
  & J^\mu = n u^\mu + \dots\,,
\end{align}
\end{subequations}
where $\epsilon \equiv -p+T\partial p/\partial T + \mu \partial p/\partial\mu$ is the energy density,  $n \equiv \partial p/\partial\mu$ is the charge density, and again the dots denote the terms of order $O(\partial)$ and higher. The conservation laws (\ref{eq:TJ-cons-1}) are satisfied identically, simply because the above $T^{\mu\nu}$ and $J^\mu$ were obtained from a gauge- and diffeomorphism-invariant generating functional.

Let us specialize to 3+1 dimensions for definiteness. Then at order $O(\partial)$, there are no invariants that could appear in the generating functional. At order $O(\partial^2)$, we write the generating functional as 
\begin{equation}
\label{eq:W2}
  W[g,A] = \int\!\!d^{4}x\,\sqrt{-g}\left[ p(T,\mu) 
  + \sum_n f_n(T,\mu) s_n^{(2)} \right] +\dots\,,
\end{equation}
where the dots denote the terms of order $O(\partial^3)$ and higher. The coefficients $f_n(T,\mu)$ are the second-order thermodynamic susceptibilities (sometimes called thermodynamic transport coefficients) which need to be determined from the microscopic theory, just like the pressure. 
Finally, $s_n^{(2)}$ are the two-derivative invariants made out of the metric, the gauge field, and the quantities in (\ref{eq:uTmu}), such as $\nabla^2 T$, $a_\mu a^\mu$, $R$, $F_{\mu\nu}F^{\mu\nu}$ etc. The invariants must be such that they do not vanish in equilibrium. We will find it convenient to write the invariants in terms of the magnetic field $B^\mu \equiv \coeff12 \epsilon^{\mu\nu\alpha\beta} u_\nu F_{\alpha\beta}$ and the vorticity vector $\Omega^\mu \equiv \epsilon^{\mu\nu\alpha\beta} u_\nu \nabla_{\!\alpha} u_\beta$. [Convention: $\epsilon^{\mu\nu\rho\sigma} = \varepsilon^{\mu\nu\rho\sigma}/\sqrt{-g}$, $\varepsilon^{0123}=1$.] The covariant versions of the flat-space identities $\partial_i B_i=0$ and $\partial_i \Omega_i = 0$ are
\begin{subequations}
\label{eq:BID-1}
\begin{align}
\label{eq:BID-1a}
  &   \nabla{\cdot}B - B{\cdot}a + E{\cdot}\Omega = 0\,,\\
\label{eq:BID-1b}
  & \nabla{\cdot}\Omega - 2 \Omega{\cdot}a = 0\,.
\end{align}
\end{subequations}
These are also true out of equilibrium. The vorticity tensor $\omega^{\mu\nu}\equiv \frac12 \Delta^{\mu\alpha}\Delta^{\nu\beta}(\nabla_{\!\alpha}u_\beta - \nabla_{\!\beta}u_\alpha)$ is related to the vorticity vector by $\omega^{\mu\nu} = -\frac12 \epsilon^{\mu\nu\rho\sigma}u_\rho \Omega_\sigma$, so that $\omega_{\mu\nu}\omega^{\mu\nu} = \frac12 \Omega^2$.

Not all invariants are independent: for example, (\ref{eq:BID-1b}) shows that the $\Omega{\cdot}a$ term in the generating functional may be absorbed into the $E{\cdot}\Omega$ term after an integration by parts and a redefinition of the $f_n$ coefficients.%
\footnote{
  For uncharged matter, the $\Omega{\cdot}a$ term in the generating functional only gives a boundary contribution.
}
Similarly, (\ref{eq:BID-1a}) shows that the $E{\cdot}\Omega$ term in the generating functional may be absorbed into the $B{\cdot}a$ and $B{\cdot}E$ terms after an integration by parts and a redefinition of the $f_n$ coefficients.
Further, in equilibrium we have
\begin{equation}
  \nabla{\cdot}a = u^\mu R_{\mu\nu}u^\nu - \coeff12 \Omega^2\,,
\end{equation}
where $R_{\mu\nu}$ is the Ricci tensor. As a result, the $u^\mu R_{\mu\nu}u^\nu$ term in the generating functional may be absorbed into the $\Omega^2$, $a^2$, and $E{\cdot}a$ terms after an integration by parts and a redefinition of the $f_n$ coefficients.
The independent second-order invariants in the generating functional were classified in \cite{Banerjee:2012iz}, using a dimensionally reduced formulation. There are seven independent invariants in a parity-preserving theory, and nine independent invariants in a parity-violating theory. If the matter degrees of freedom do not couple to the gauge field, there are only three independent invariants.
We choose the independent invariants as listed in Table~\ref{tab:T2}. This fixes the definition of the susceptibility coefficients $f_n(T,\mu)$ in the generating functional~(\ref{eq:W2}).

\begin{table}
\begin{center}
\def\arraystretch{1.2}
\setlength\tabcolsep{4pt}
\begin{tabular}{|c|c|c|c|c|c|c|c|c|c|}
 \hline
 \hline
 $n$ & 1 & 2 & 3 & 4 & 5 & 6 & 7 & 8 & 9\\ 
 \hline
 \hline
 $s^{(2)}_n$
 & $R$   % 
 & $a^2$  %
 & $\Omega^2$
 & $B^2$ 
 & $B{\cdot}\Omega$
 & $E^2$
 & $E{\cdot}a$
 & $B{\cdot}E$
 & $B{\cdot}a$
 \\
 \hline
  P  & $+$ & $+$ & $+$ & $+$ & $+$ & $+$ & $+$ & $-$ & $-$\\
  \hline
  C  & $+$ & $+$ & $+$ & $+$ & $-$ & $+$ & $-$ & $+$ & $-$\\
  \hline
  T  & $+$ & $+$ & $+$ & $+$ & $+$ & $+$ & $+$ & $-$ & $-$\\
  \hline
  W  & n/a  & n/a  & 2  & 4  & 3  & 4  & n/a  & 4  & n/a\\
  \hline
\end{tabular}
\end{center}
\caption{Independent $O(\partial^2)$ equilibrium invariants in 3+1 dimensions. The rows labeled P, C, T indicate the eigenvalue of the corresponding invariant under parity, charge conjugation, and time-reversal, respectively. The row labeled W indicates the conformal weight $w$ of the corresponding invariant. The invariants labeled ``n/a'' do not transform homogeneously under the Weyl rescaling of the metric. The first invariant is the Ricci scalar, the other invariants are formed out of the vectors defined in the text.}
\label{tab:T2}
\end{table}

The table also lists the conformal weights $w$ of the invariants under the Weyl rescaling of the metric $g_{\mu\nu}\to \tilde g_{\mu\nu} = e^{-2\varphi} g_{\mu\nu}$. The quantity $\Phi$ has conformal weight $w$ if under the Weyl rescaling $\Phi \to \tilde \Phi = e^{w\varphi} \Phi$. The zeroth-order invariants $T$ and $\mu$ have $w=1$. The acceleration transforms inhomogeneously, $\tilde a_\mu = a_\mu -\partial_\mu \varphi$. In a conformal theory, the generating functional $W[g,A]$ must be invariant under the Weyl rescaling. While the invariants $R$, $a^2$, and $E{\cdot}a$ do not have well-defined weights, in $d+1$ spacetime dimensions the combination
\begin{equation}
  \int\!\!d^{d+1}x\,\sqrt{-g}\, \Big( f (R + d(d{-}1) a^2) - 2d(\partial f/\partial\mu)E{\cdot}a\Big)
\end{equation}
is Weyl-invariant up to a boundary term, for $f(T,\mu) = T^{d-1}F(\mu/T)$.
Thus in a conformal theory in 3+1 dimensions, $f_1 = T^2 F(\mu/T)$, $f_2 = 6 f_1$, $f_7 = -6 \partial f_1/\partial\mu$, and $f_9 = 0$. In particular, thermal equilibrium of a neutral conformal fluid in 3+1 dimensions is characterized by two independent second-order susceptibility coefficients $f_1$ and $f_3$.

\subsection{Trace anomaly}
\label{SS:TA}
In a conformal theory, quantum effects give rise to the conformal anomaly~\cite{Duff:1993wm}. For a conformal field theory subject to external gravitational and electromagnetic fields, the trace of the energy-momentum tensor is
\begin{equation}
\label{eq:trace-anomaly}
  g_{\mu\nu}T^{\mu\nu} = -\frac{a}{16\pi^2}\left( R_{\mu\nu\rho\sigma}^2 - 4R_{\mu\nu}^2 + R^2\right) 
  + \frac{c}{16\pi^2} \left( R_{\mu\nu\rho\sigma}^2 - 2R_{\mu\nu}^2 +\coeff13 R^2\right)
  -\frac{b_0}{4} F_{\mu\nu}^2\,.
\end{equation}
Here $a$ and $c$ are dimensionless coefficients that depend on the degrees of freedom of the theory. For example, for a free theory of $N_S$ real scalars, $N_F$ Dirac fermions, and $N_V$ vector fields, one has
\begin{align*}
   a = \frac{1}{360} \left( N_S + 11 N_F + 62 N_V \right)\,,\ \ \ \ 
   c = \frac{1}{120} \left( N_S + 6 N_F + 12 N_V \right)\,. 
\end{align*}
The coefficient $b_0$ is the coefficient of the leading-order beta function for the electromagnetic coupling $e$ used to couple the theory to the external gauge field: $M \frac{d}{dM}(1/e^2) = -b_0 + O(e^2)$, where $M$ is the renormalization scale%
\footnote{The external electromagnetic field $A_\mu$ can be introduced by minimally coupling the fundamental matter fields of the theory to $A_\mu$ (without factors of $e$) and adding the kinetic term $-\frac{1}{4e^2}F_{\mu\nu}^2$ to the action of the theory. The electromagnetic field becomes non-dynamical as $e\to0$. The last term in (\ref{eq:trace-anomaly}) describes the violation of scale invariance due to the renormalization-group running of $e$, caused by the charged matter fields of the theory. We have found the discussion in Section~2 of~\cite{Fuini:2015hba} helpful.
}. 
For example, for a free theory of $n_s$ complex scalars with charges $q_{s,k}$ ($k=1,\dots,n_s$) and $n_f$ Dirac fermions with charges $q_{f,i}$ ($i=1,\dots,n_f$), one has
$$
  b_0 = \frac{1}{6\pi^2}\left( \sum_{i=1}^{n_f} q_{f,i}^2 + \coeff14 \sum_{k=1}^{n_s}q_{s,k}^2 \right)\,,
$$
which gives the standard one-loop QED beta-function.

The gravitational contributions to the trace anomaly in (\ref{eq:trace-anomaly}) are fourth order in derivatives. As we are only interested in the generating functional (and hence the energy-momentum tensor) up to second order in derivatives, we will ignore these contributions (see \cite{Eling:2013bj} for a general discussion in an arbitrary number of dimensions). The electromagnetic contribution in (\ref{eq:trace-anomaly}) is, however, second order in derivatives, and must emerge from the generating functional. This can be accounted for if the coefficients $f_4$ and $f_6$ are not themselves Weyl-invariant. In fact, the trace anomaly places constraints on the form of these coefficients.

Indeed, consider the ``4,6'' part of the generating functional, $W_{4,6} = \int\!\sqrt{-g}\,(f_4 B^2 + f_6 E^2)$. For the corresponding energy-momentum tensor $T^{\mu\nu}_{4,6}$ one finds
$$
  g_{\mu\nu}T^{\mu\nu}_{4,6} = -f_4' B^2 - f_6' E^2\,,
$$
where $f_n' \equiv Tf_{n,T} + \mu f_{n,\mu}$, and the comma denotes the derivative with respect to the argument that follows. For the trace anomaly, this has to match $-b_0\frac{1}{4} F_{\mu\nu}^2 = b_0\frac{1}{2} (E^2 - B^2)$, which gives $f_4'=-f_6'=\frac{b_0}{2}$. This is solved by 
\begin{equation}
\label{eq:f4f6}
  f_4 = \frac{b_0}{2} \ln\frac{T}{M} + C_4(\mu/T)\,,\ \ \ \ 
  f_6 = -\frac{b_0}{2} \ln\frac{T}{M} + C_6(\mu/T)\,,
\end{equation}
where the integration constant $M$ can be interpreted as the renormalization scale. This explicitly shows that $f_4$ and $f_6$ shift under the Weyl rescaling, due to the $\ln T$ terms. The total ``electromagnetic'' part of the generating functional is then
\begin{align*}
  W_{EM} & \equiv \int\!\sqrt{-g}\, \left(f_4 B^2 + f_6 E^2 - \frac{1}{4e^2}F_{\mu\nu}^2\right) \\
         & = -\frac{1}{4}\int\!\sqrt{-g} \left[ \frac{1}{e^2(M)} + b_0 \ln \frac{M}{T} \right]F_{\mu\nu}^2 
         + \int\!\sqrt{-g} \left(F_4 B^2 + F_6 E^2 \right)\,,
\end{align*}
where $e^2(M)$ is the renormalized coupling, and $F_4(\mu/T)$, $F_6(\mu/T)$ are the renormalized susceptibilities. The renormalization-group equation for $1/e^2(M)$ ensures that $M \frac{d}{dM} W_{EM} = 0$, i.e.\ the generating functional does not depend upon the renormalization scale, as the case should be.

\subsection{The energy-momentum tensor and the current}
\label{SS:EMTC}
Let us now write down the energy-momentum tensor that follows from the generating functional (\ref{eq:W2}). This was done in \cite{Banerjee:2012iz, Jensen:2012jh} for neutral matter, and in \cite{Bhattacharyya:2014bha} for charged matter, in a dimensionally reduced formulation. Here we will write the energy-momentum tensor in the covariant form, decomposing
$T^{\mu\nu}$ with respect to the fluid velocity $u^\mu$ as
\begin{equation}
\label{eq:Tmn}
  T^{\mu\nu} = {\cal E} u^\mu u^\nu + {\cal P}\Delta^{\mu\nu} +{\cal Q}^\mu u^\nu + {\cal Q}^\nu u^\mu + {\cal T}^{\mu\nu}\,.
\end{equation}
The energy density is ${\cal E}\equiv u_\mu T^{\mu\nu}u_\nu$, the pressure is ${\cal P}\equiv \frac13 \Delta_{\mu\nu}T^{\mu\nu}$, the energy flux ${\cal Q}_\mu \equiv - \Delta_{\mu\alpha} T^{\alpha\beta}u_\beta$ is transverse to $u^\mu$, and the stress ${\cal T}^{\mu\nu} \equiv T^{\langle \mu\nu\rangle}$ is transverse to $u_\mu$, symmetric, and traceless. The angular brackets denote the symmetric transverse traceless part of a tensor, $X_{\langle \mu\nu\rangle} \equiv \frac12 (\Delta_{\mu\alpha} \Delta_{\nu\beta} + \Delta_{\nu\alpha} \Delta_{\mu\beta} -\frac23 \Delta_{\mu\nu} \Delta_{\alpha\beta}) X^{\alpha\beta}$.
Similarly, the $U(1)$ current can be written as 
\begin{equation}
\label{eq:Jmu}
  J^\mu = {\cal N} u^\mu + {\cal J}^\mu\,,
\end{equation}
where ${\cal N}\equiv -u_\mu J^\mu$ is the charge density, and ${\cal J}^\mu \equiv \Delta^\mu_\nu J^\nu$ is the spatial current. Comparing with (\ref{eq:JM}), we find~\cite{Kovtun:2016lfw}%
\begin{subequations}
\label{eq:NJ-M}
\begin{align}
  & {\cal N} = \rho - \nabla{\cdot}p + p{\cdot}a - m{\cdot}\Omega\,,\\[5pt]
  & {\cal J}^\mu = \epsilon^{\mu\nu\rho\sigma}u_\nu \left( \nabla_{\!\rho} + a_\rho \right) m_\sigma\,,
\end{align}
\end{subequations}
where $p^\mu \equiv u_\nu M^{\nu\mu}$ is the electric polarization vector, $m^\mu \equiv \frac12 \epsilon^{\mu\nu\rho\sigma} u_\nu M_{\rho\sigma}$ is the magnetic polarization vector, and $\rho= \partial{\cal F}/\partial\mu$ is the density of free charges.

As an example, consider matter that has a global $U(1)$ charge (so that one can introduce the corresponding chemical potential), but which is not subject to any external electric and magnetic fields coupled to that $U(1)$ current. An example is QCD at finite (or zero) baryon number chemical potential. 
A straightforward (and tedious) calculation gives the coefficients of the energy-momentum tensor (\ref{eq:Tmn}) in terms of the three susceptibilities $f_n(T,\mu)$ as%
\begin{subequations}
\label{eq:cr-statics}
\begin{align}
\label{eq:EE}
  & {\cal E}  =  \epsilon + (f_1' - f_1) R + (4 f_1' + 2f_1'' - f_2 - f_2')a^2 \nonumber \\
  & \ \ \ \ \ \, + (f_1' - f_2 - 3f_3 + f_3')\, \Omega^2 - 2(f_1 + f_1' - f_2)\, u^\alpha R_{\alpha\beta} u^\beta \,,\\[5pt]
   & {\cal P}  =  p + \coeff13 f_1 R - \coeff13 (2f_1' +f_3)\, \Omega^2 -\coeff13 (2f_1' + 4f_1'' - f_2)a^2 +\coeff23 (2f_1' - f_1)\, u^\alpha R_{\alpha\beta} u^\beta\,,\\[5pt]
\label{eq:QQ}
   & {\cal Q}_\mu  =  (f_1' + 2f_3')\, \epsilon_{\mu\lambda\rho\sigma}a^\lambda u^\rho \Omega^\sigma + (2f_1 + 4f_3) \Delta_\mu^\rho R_{\rho\sigma} u^\sigma\,,\\[5pt]
   & {\cal T}_{\mu\nu}  =  (4f_1' + 2f_1'' - 2f_2) a_{\langle\mu} a_{\nu\rangle} -\coeff12 (f_1' -4f_3)\, \Omega_{\langle\mu} \Omega_{\nu\rangle} + 2f_1'\, u^\alpha R_{\alpha\langle \mu\nu \rangle\beta} u^\beta -2f_1 R_{\langle \mu\nu\rangle}\, ,
\end{align}
\end{subequations}
where again 
\begin{align*}
f_n'&\equiv  T f_{n,T} + \mu f_{n,\mu}\, ,\\
f_n''&\equiv  T^2 f_{n,T,T} + 2\mu T f_{n,T,\mu} + \mu^2 f_{n,\mu,\mu}\, ,
\end{align*}
and the comma subscript denotes the partial derivative with respect to the argument that follows. The leading-order energy density is $\epsilon = -p + Tp_{,T} + \mu p_{,\mu}$, as before. Equations (\ref{eq:Tmn}) and (\ref{eq:cr-statics}) give the energy-momentum tensor of a relativistic fluid in hydrostatic equilibrium, up to $O(\partial^2)$ terms beyond the perfect fluid approximation. This generalizes the result of~\cite{Jensen:2012jh} to non-zero~$\mu$.

The polarization vectors which determine the equilibrium current (\ref{eq:NJ-M}) are
\begin{subequations}
\label{eq:pm-vectors}
\begin{align}
  & p^\alpha = 2f_6\, E^\alpha + f_7\, a^\alpha + f_8\, B^\alpha\,,\\[5pt]
  & m^\alpha = 2f_4\, B^\alpha + f_5\, \Omega^\alpha + f_8\, E^\alpha + f_9\, a^\alpha\,.
\end{align}
\end{subequations}
As an example, consider parity-invariant matter that has a global $U(1)$ charge (so that one can introduce the corresponding chemical potential), but which is not subject to any external electric and magnetic fields coupled to that $U(1)$ current.  Again, QCD at finite (or zero) baryon number chemical potential would be an example. The charge density and the spatial current in (\ref{eq:Jmu}) are then
\begin{subequations}
\label{eq:NJ-cr-statics}
\begin{align}
\label{eq:NN}
  & {\cal N} = n + f_{1,\mu}R + (f_{2,\mu} + f_7 + f_7') a^2 + \left( f_{3,\mu} - f_5 + \coeff12 f_7 \right) \Omega^2 - f_7\, u^\alpha R_{\alpha\beta}u^\beta\,,\\[5pt]
  & {\cal J}^\mu = -(f_5 + f_5')\epsilon^{\mu\nu\rho\sigma}u_\nu a_\rho \Omega_\sigma +2f_5\Delta^{\mu\rho}R_{\rho\lambda} u^\lambda\,,
\end{align}
\end{subequations}
where $n\equiv\partial p/\partial\mu$ is the zeroth-order charge density.

We see that the thermodynamics of QCD with a baryon number chemical potential is specified by the pressure $p(T,\mu)$ at zeroth order in derivatives, as well as by the five susceptibilities $f_i(T,\mu)$, with $i=1,2,3,5,7$ at second order in derivatives. For a conformal theory, $f_2 = 6f_1$, $f_7 = -6 \partial f_1/\partial\mu$, $f_9 = 0$, and hence one only needs three susceptibilities $f_1$, $f_3$, and~$f_5$ to specify the two-derivative thermodynamics of conformal matter not subject to external electromagnetic fields.

\subsection{Kubo formulas}
\label{SS:KF}
The above equilibrium expressions for $T^{\mu\nu}[g]$ and $J^\mu[g,A]$ allow for a straightforward computation of equilibrium (zero frequency) correlation functions of the corresponding operators. In order to write down the correlation functions for matter at rest in flat space, we choose $V^\alpha = (1,{\bf 0})$, and take the external sources as $g_{\mu\nu} = \eta_{\mu\nu} + \delta g_{\mu\nu}(\k) e^{i\k{\cdot}\x}$, $A_\lambda = \mu_0 \delta_\lambda^0 + \delta A_\lambda(\k) e^{i\k{\cdot}\x}$. The equilibrium two-point functions $G_{AB}$ of two operators $A$ and $B$ are then defined by varying the corresponding equilibrium one-point functions with respect to the source,
\begin{subequations}
\label{eq:GGG-def}
\begin{align}
  &   \delta_g(\sqrt{-g}\, T^{\mu\nu}) 
  = \coeff12 G_{T^{\mu\nu} T^{\alpha\beta}}(\omega{=}0, \k)\, \delta g_{\alpha\beta} (\k)\,, \\[5pt]
  &  \delta_g(\sqrt{-g}\, J^{\mu}) 
  = \coeff12 G_{J^{\mu} T^{\alpha\beta}}(\omega{=}0, \k)\, \delta g_{\alpha\beta} (\k)\,,\\[5pt]
  & \delta_A(\sqrt{-g}\, J^{\mu}) 
  = G_{J^{\mu} J^{\nu}}(\omega{=}0, \k)\, \delta A_{\nu} (\k)\,.
\end{align}
\end{subequations}
The locality of the derivative expansion implies that the two-point functions are at most quadratic in $\k$, with the coefficients of the $O(\k^2)$ terms determined by the susceptibilities~$f_n$. The energy-momentum tensor (\ref{eq:Tmn}), (\ref{eq:cr-statics}) implies the following Kubo formulas in terms of the above zero-frequency correlation functions:
\begin{align}
\label{eq:f1-Kubo}
  & f_1 = - \coeff12 \lim_{\k\to0} \frac{\partial^2}{\partial k_z^2} G_{T^{xy} T^{xy}} 
        = \coeff14 \lim_{\k\to0} \frac{\partial^2}{\partial k_z^2} G_{T^{xx} T^{yy}} \,,\\[5pt]
\label{eq:f2-Kubo}
  & f_2 = \coeff14 \lim_{\k\to0} \frac{\partial^2}{\partial k_z^2} \left( G_{T^{tt} T^{tt}} + 2 G_{T^{tt} T^{xx}} - 4 G_{T^{xy} T^{xy}}\right) \,,\\[5pt]
\label{eq:f3-Kubo}
  & f_3 = \coeff14 \lim_{\k\to0} \frac{\partial^2}{\partial k_z^2} \left( G_{T^{tx} T^{tx}} + G_{T^{xy} T^{xy}} \right) \,.
\end{align}
There are of course other ways to write the Kubo formulas for $f_{1,2,3}$ which follow from the rotation invariance of the two-point functions.%
\footnote{
For example, $(2f_1 {-} T f_{1,T} {-} \mu f_{1,\mu}) = \frac14 \lim_{\k\to0} \frac{\partial^2}{\partial k_z^2} G_{T^\mu_{\ \mu} T^{xx}}$, $(6f_1 {-} f_2) = \frac14 \lim_{\k\to0} \frac{\partial^2}{\partial k_z^2} (G_{T^\mu_{\ \mu} T^{tt}} +4 G_{T^\mu_{\ \mu} T^{xx}})$. As expected, in a CFT with $T^\mu_{\ \,\mu}=0$ one recovers the constraints $f_1 = T^2 F(\mu/T)$, $f_2 = 6 f_1$.
}
Similarly, the equilibrium current (\ref{eq:Jmu}), (\ref{eq:NJ-M}), (\ref{eq:pm-vectors}) gives the following Kubo formulas
\begin{align}
  & f_4 = \coeff14 \lim_{\k\to0} \frac{\partial^2}{\partial k_z^2} G_{J^{x} J^{x}}\,,\\[5pt]
  & f_5 = \coeff12 \lim_{\k\to0} \frac{\partial^2}{\partial k_z^2} G_{J^{x} T^{tx}}\,,\\[5pt]
  & f_6 = \coeff14 \lim_{\k\to0} \frac{\partial^2}{\partial k_z^2} G_{J^{t} J^{t}}\,,\\[5pt]
\label{eq:f7-Kubo}
  & f_7 = -\coeff12 \lim_{\k\to0} \frac{\partial^2}{\partial k_z^2} (G_{J^{t} T^{tt}} + G_{J^{t} T^{xx}})\,.
\end{align}
Thus all seven parity-preserving thermodynamic susceptibilities admit Kubo formulas in terms of equilibrium two-point functions in flat space without external fields. The parity-breaking susceptibilities $f_8$ and $f_9$ do not appear in the above linearized analysis, but can be expressed in terms of equilibrium three-point functions in flat space without external fields.%
\footnote{
One can write down Kubo formulae for parity-breaking thermodynamic transport coefficients in 2+1 dimensions in terms of equilibrium two-point functions, see \cite{Jensen:2011xb}. }

In order to find the three-point functions, we expand the equilibrium $T^{\mu\nu}$ and $J^\mu$ to quadratic order in small fluctuations $h_{\alpha\beta}(\x)$, $A_\lambda(\x)$. Note that we don't need to solve for the conservation of $T^{\mu\nu}$ in our setup -- the conservation laws (\ref{eq:TJ-cons-1}) are satisfied {\em identically} in equilibrium, as a consequence of the diffeomorphism invariance of $W[g,A]$, and $V$ being a Killing vector. For example, let us take $h_{tt}(z)$ and $h_{tx}(y)$ as the only non-vanishing perturbations. We then find
\begin{align}
\label{eq:f9-kubo}
  J^t = p_{,\mu} + O(h_{tt}, h_{tt}'') - \coeff12 f_9 h_{tt}'(z) h_{tx}'(y) + O\left( h_{tx} h_{tx}'', h_{tx}'^2 , h_{tt} h_{tt}'', h_{tt}'^2, h_{tt}^2 \right)\,.
\end{align}
As another example, let us take $h_{tt}(z)$ and $A_{x}(y)$ as the only non-vanishing perturbations. We then find
\begin{align}
\label{eq:f89-kubo}
  J^t = p_{,\mu} + O(h_{tt}, h_{tt}'') + \coeff12 h_{tt}'(z) A_x'(y) (f_8' + f_{9,\mu}) + O\left( A_x'^2 , h_{tt} h_{tt}'', h_{tt}'^2, h_{tt}^2 \right)\,.
\end{align}
Taking the variation of the one-point functions (\ref{eq:f9-kubo}), (\ref{eq:f89-kubo}) with respect to the sources $h_{\alpha\beta}$, $A_\alpha$, one finds that the susceptibilities $f_9$ and $f_8' + f_{9,\mu}$ are given in terms of the second derivatives of the appropriately defined three-point functions $G_{J^t T^{tx} T^{tt}}(p,k)$ and $G_{J^t J^x T^{tt}}(p,k)$, respectively.

\section{Free fields}
\begin{figure}
\centering
\includegraphics[width=0.25\textwidth]{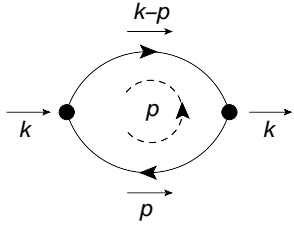}
\caption{The one-loop diagram contributing to the $\langle \hat T^{\mu\nu}(\k) \hat T^{\rho\sigma}(-\k) \rangle$ correlation function for fermions. The two vertices denote the two energy-momentum tensor insertions, the lines are the fundamental field propagators, and $p$ denotes the loop momentum which needs to be integrated out. The hat in $\hat T^{\mu\nu}$ signifies that it is an operator made out of the fundamental fields. Similar diagrams arise for free scalar as well as for free gauge fields.}
\label{fig:1-loop}
\end{figure}

Let us now use the above Kubo formulas to evaluate the thermodynamic susceptibilities for non-interacting quantum fields in 3+1 dimensions. The energy-momentum tensor and the current are quadratic in the fields, hence the two-point functions can be evaluated from the diagram schematically shown in Figure~\ref{fig:1-loop}.%
The diagram can be evaluated by the standard methods of equilibrium thermal field theory in flat space~\cite{Bellac:2011kqa, Kapusta:2006pm}. The integration over the intermediate momenta in the loop will give rise to ultraviolet divergences which can be regulated by introducing a high-momentum cutoff scale~$\Lambda$. We will assume that the cutoff dependence is removed by the standard zero-temperature renormalization, and will only report the temperature-dependent (and cutoff-independent) contributions to the thermodynamic susceptibilities~$f_n$.
[As an example, the zero-temperature contribution to $f_1$ (which we will not indicate explicitly) gives rise to the renormalization of Newton's constant by the quantum fluctuations of the matter fields. Similarly, the zero-temperature contributions to $f_4$ and $f_6$ give rise to electric charge renormalization.]

\subsection{Scalars}
\label{SS:Scalars}
We start with the massless real scalar field. The action is~\cite{Parker:2009uva}
$$
S = - \coeff12 \int\! d^4x \sqrt{-g}\left( g^{\mu\nu} \partial_\mu \phi\, \partial_\nu \phi + \xi R \phi^2 \right),
$$
where $R$ is the Ricci scalar as before, and the dimensionless parameter $\xi$ specifies the coupling to curvature. The scalar field is minimally coupled for $\xi = 0$, and conformally coupled for $\xi =1/6$. 
The energy-momentum tensor of the theory is found by varying the action with respect to the metric,
$$
\hat T^{\mu\nu} = \nabla^{\mu} \phi \, \nabla^{\nu} \phi - \coeff12\, g^{\mu\nu} g^{\alpha\beta} \partial_\alpha \phi \, \partial_\beta \phi - \xi \left(\nabla^\mu \nabla^\nu - g^{\mu\nu} \nabla^2\right)\phi^2 + \xi \phi^2 G^{\mu\nu},
$$
where $G^{\mu\nu} = R^{\mu\nu} - \coeff12 g^{\mu\nu} R$ is the Einstein tensor. We use the hat to distinguish the microscopic $\hat T^{\mu\nu}$ (which depends on the fundamental fields) from the macroscopic $T^{\mu\nu}$ defined by~(\ref{eq:TJ-def}) (which only depends on the temperature, chemical potential, and the fixed external sources). The explicit metric dependence in the $\hat T^{\mu\nu}$ operator will give rise to contact terms in two-point functions upon taking the metric variation. Schematically,
$$
  \hat T^{\mu\nu} = O\!\left(g (\partial\phi)^2\right) +O\!\left( g\, \partial^2 \phi^2\right) + O\!\left(\partial g\partial\phi^2\right) + O\!\left(\phi^2\partial^2 g\right)\,.
$$
We will be computing two-point correlation functions of $\hat T^{\mu\nu}$ in flat space, by taking the metric as $g_{\mu\nu} = \eta_{\mu\nu} + \delta g_{\mu\nu}$. In this case $\langle\phi^2\rangle$ does not vary in space, and $\partial\langle\phi^2\rangle$ vanishes. The term $\langle (\partial\phi)^2 \rangle$ contributes to the internal energy of the scalar field in equilibrium, and gives a constant momentum-independent contribution to the two-point function. Thus for the purpose of computing the thermodynamic susceptibilities (which appear as $O(k^2)$ contributions to the two-point function), the only relevant contact term arises from $\langle \phi^2\rangle \partial^2 g$.
The variation can be written as
$$
  \frac{\delta}{\delta g_{\alpha\beta}(y)} \sqrt{-g} \;\hat T^{\mu\nu}(x) = 
  A^{\mu\nu,\alpha\beta} \delta(x-y) + B^{\mu\nu,\alpha\beta,\rho} \partial_\rho \delta(x-y) + 
  C^{\mu\nu,\alpha\beta,\rho\sigma} \partial_\rho \partial_\sigma \delta(x-y)\,,
$$
with the coefficients $A,B,C$ that are local functions of the field $\phi$. Expanding the Einstein tensor, we find $C^{\mu\nu,\alpha\beta,\rho\sigma} = \frac12 \xi \phi^2 P^{{\mu\nu,\alpha\beta,\rho\sigma}}$, with
\begin{align*}
  P^{\mu\nu,\alpha\beta,\rho\sigma} & =   \eta^{\mu(\alpha} \eta^{\beta) (\sigma} \eta^{\rho)\nu} + \eta^{\mu (\rho} \eta^{\sigma) (\beta} \eta^{\alpha) \nu} - \eta^{\mu (\alpha} \eta^{\beta) \nu} \eta^{\rho\sigma}  \\[5pt]
 & -  \eta^{\mu (\rho} \eta^{\sigma) \nu} \eta^{\alpha\beta} - \eta^{\mu\nu}\eta^{\alpha (\rho} \eta^{\sigma) \beta} + \eta^{\mu\nu}\eta^{\rho\sigma}\eta^{\alpha\beta} \,,
\end{align*}
where the parentheses denote symmetrization (with the 1/2). Note that $P^{\mu\nu,\alpha\beta,\rho\sigma} = P^{\alpha\beta,\mu\nu,\rho\sigma}$. The ``variational'' two-point function $G_{T^{\mu\nu} T^{\alpha\beta}}$ defined by (\ref{eq:GGG-def}) is then related to the standard ``operator'' two-point function $\langle \hat T^{\mu\nu} \hat T^{\alpha\beta} \rangle$ by%
\footnote{
	To see how this relation arises, one can start with the Euclidean functional integral representation, with the action $iS[g]_{t\to-i\tau, g_{00}\to-g_{00}^\E, g_{0k}\to i g_{0k}^\E, g_{kl}\to g_{kl}^\E} = -S_\E[g^\E]$. The Euclidean energy-momentum tensor is defined by $\delta S_\E = -\frac12 \int\! \sqrt{g_\E}\, T^{\mu\nu}_\E \delta g_{\mu\nu}^\E$, so that $\delta g_{\mu\nu}T^{\mu\nu} = \delta g_{\mu\nu}^\E T^{\mu\nu}_\E$, e.g.\ $T^{00}|_{t\to-i\tau, g_{00}\to-g_{00}^\E, g_{0k}\to i g_{0k}^\E, g_{kl}\to g_{kl}^\E} = -T^{00}_\E$.
}
\begin{equation}
\label{eq:GTT-contact}
  G_{T^{\mu\nu} T^{\alpha\beta}}(\k) = \langle \hat T^{\mu\nu} \hat T^{\alpha\beta} \rangle(\k) - \xi \langle\phi^2\rangle P^{\mu\nu,\alpha\beta,\rho\sigma} k_\rho k_\sigma\,.
\end{equation}
The terms in the right-hand side of (\ref{eq:GTT-contact}) may be evaluated diagrammatically by the standard rules of equilibrium thermal field theory in flat space in the Matsubara formalism. As the real field is uncharged, the chemical potential $\mu$ is not relevant. The Euclidean propagator is $D(i\omega_n,\k) = [-(i\omega_n)^2 + \k^2]^{-1}$, where $\omega_n = 2\pi nT$, with integer~$n$.

The contact term contributes a simple ``bubble'' diagram with
$$
  \langle \phi^2\rangle = \frac{T^2}{12}\,.
$$
The susceptibilities $f_1$, $f_2$, $f_3$ may be computed from the Kubo formulas (\ref{eq:f1-Kubo}), (\ref{eq:f2-Kubo}), (\ref{eq:f3-Kubo}), using the diagrams schematically shown in Fig.~\ref{fig:1-loop}. Performing the integral over the intermediate momenta, we find for the temperature-dependent contributions
\begin{align}
   f_1 = \frac{T^2}{144} \left( 1-6\xi \right)\,,\ \ \ \ 
   f_2 = 0\,,\ \ \ \ 
   f_3 = -\frac{T^2}{144}\,.
\end{align}
The rest of the susceptibilities $f_n$ vanish for the real scalar field. 

For a complex scalar field at $\mu=0$, the above $f_{1,2,3}$ get multiplied by a factor of~2. Minimally coupling the complex scalar field to the external gauge field $A_\mu$ gives
\begin{equation}
  f_4 = -f_6 = \frac{1}{48\pi^2} \ln \frac{T}{M}\,,
\end{equation}
according to the general result~(\ref{eq:f4f6}). The rest of the susceptibilities $f_n$ vanish at $\mu=0$ by charge conjugation and parity.

\subsection{Dirac fermions}
\label{SS:Fermions}
We now consider a massless Dirac fermion field at $\mu = 0$. The action is given by~\cite{Parker:2009uva}
$$
S = - \, i \int\! d^4x \sqrt{-g} \; \bar{\Psi} \underline{\gamma}^\mu \nabla_\mu \Psi.
$$
Here $\underline{\gamma}^\mu$ are the spacetime dependent Dirac $\gamma$-matrices,
$
\underline{\gamma}^\mu (x) = e^\mu_a (x) \, \gamma^a,
$
where $e^\mu_a(x)$ is the vierbein field, and $\gamma^a$, $a = 0,1,2,3$ are the usual position independent flat space $\gamma$-matrices. The Clifford algebras satisfied by the $\underline{\gamma}$ and $\gamma$ matrices are
\begin{align*}
 \left\{ \underline{\gamma}^\mu(x), \underline{\gamma}^\nu(x) \right\} = 2 \, g^{\mu\nu}(x),\ \ \ \ 
 \left\{ \gamma^a, \gamma^b \right\} = 2\, \eta^{ab}.
\end{align*}
The covariant derivative acting on the fermion field is given by
$$
\nabla_{\!\mu} \Psi = \partial_\mu \Psi + \coeff12 \,\omega_\mu^{ab} \,\sigma_{ab}\, \Psi,
$$
where $\sigma_{ab} \equiv \frac14 [ \gamma_a,\gamma_b ]$, and $\omega_\mu^{ab}$ is the spin connection,
$$
\omega_\mu^{ab} = \coeff12 \, e^{a\nu} \left(\partial_\mu e^b_\nu - \partial_\nu e^b_\mu \right) - \coeff12\, e^{b\nu} \left(\partial_\mu e^a_\nu - \partial_\nu e^a_\mu \right) + \coeff12 \, e^{a\nu} e^{b\rho} \left(\partial_\rho e^c_\nu - \partial_\nu e^c_\rho\right) e_{c\mu}.
$$
The energy-momentum tensor is 
$$
  \hat T^{\mu\nu} = \frac{2}{\sqrt{-g}}\frac{\delta S}{\delta g_{\mu\nu}} = \frac{e^\nu_a}{\sqrt{-g}} \frac{\delta S}{\delta e_{a\mu}}\,,
$$
which gives
\begin{equation}
\label{eq:Tmn-fermions}
  \hat T^{\mu\nu} = 
  \frac{i}{4} \left( 
    \bar{\Psi} \underline{\gamma}^{\mu} \nabla^\nu \Psi 
  - \nabla^\mu \bar{\Psi} \,\underline{\gamma}^{\nu} \Psi 
  + \bar{\Psi} \underline{\gamma}^{\nu} \nabla^\mu \Psi 
  - \nabla^\nu \bar{\Psi} \, \underline{\gamma}^{\mu} \Psi 
  \right).
\end{equation}
There are no terms in the energy-momentum tensor with two derivatives of the metric, hence there are no contact terms analogous to the ones we had for the scalar field. Hence we have
\begin{equation}
\label{eq:GTT-2}
  G_{T^{\mu\nu} T^{\alpha\beta}}(\k) = \langle \hat T^{\mu\nu} \hat T^{\alpha\beta} \rangle(\k) \,,
\end{equation}
where the right-hand side may be evaluated with the flat-space energy-momentum tensor (replacing the covariant derivatives in (\ref{eq:Tmn-fermions}) with partial derivatives) by the standard rules of equilibrium thermal field theory in the Matsubara formalism. The Euclidean propagator is $D(i\omega_n,\k) = \slashed{k}[-(i\omega_n)^2 + \k^2]^{-1}$, with $\slashed{k} = \gamma^0_E \omega_n + \gamma{\cdot}\k$, with $\gamma^0_E = i \gamma^0$ and $\omega_n = (2n+1)\pi T$, with $n$ integer. We get
\begin{equation}
  f_1 = -\frac{T^2}{144}\,,\ \ \ \ 
  f_2 = -\frac{T^2}{24}\,,\ \ \ \ 
  f_3 = -\frac{T^2}{288}\,.
\end{equation}

Minimally coupling the Dirac current, $\hat{J}^\mu = - \bar{\Psi} \underline{\gamma}^\mu \Psi$, to the external gauge field $A_\mu$ gives
\begin{equation}
  f_4 = -f_6 = \frac{1}{12\pi^2} \ln \frac{T}{M}\,,
\end{equation}
according to the general result~(\ref{eq:f4f6}). The rest of the susceptibilities $f_n$ vanish at $\mu=0$ by charge conjugation and parity.

\subsection{Gauge fields}
\label{SS:Gauge}
We now give results for the thermodynamic susceptibilities of a free $U(1)$ gauge field. The action for the theory is given by
\begin{equation*}
S = - \frac{1}{4} \int d^4 x \sqrt{-g} \, F^{\mu\nu} F_{\mu\nu},
\end{equation*}
with the energy-momentum tensor 
\begin{equation*}
\hat{T}^{\mu\nu} = F^{\mu\alpha} F^{\nu}_{~\,\alpha} - \frac{1}{4} \, g^{\mu\nu} F^{\alpha\beta} F_{\alpha\beta}.
\end{equation*}
Once again there are no contact term contributions to the two-point function, and the relation (\ref{eq:GTT-2}) between the variational and operator definitions of the correlation function is valid. Proceeding in the same way as in \cite{Romatschke:2009ng} and evaluating the one-loop diagram similar to figure~\ref{fig:1-loop} by using the Euclidean propagator for the gauge field in the Feynman gauge, $D^{\mu\nu}(i\omega_n,\k) = \delta^{\mu\nu} [-(i\omega_n)^2 + \k^2]^{-1}$ with $\omega_n = 2\pi n T$, one finds for the thermodynamic susceptibilities 
\begin{equation}
  f_1 = -\frac{T^2}{36}\,,\ \ \ \ 
  f_2 = -\frac{T^2}{6}\,,\ \ \ \ 
  f_3 = \frac{T^2}{36}\, ,
\end{equation}
with the other susceptibilities vanishing.

\section{Discussion}
\label{SS:Discussion}
The emphasis of this note was on the Kubo formulas for thermodynamic susceptibilities that appear at two-derivative order in the constitutive relations of the energy-momentum tensor and of the global $U(1)$ current (such as the baryon number current in QCD). Our work is close in spirit to \cite{Jensen:2012jh, Moore:2012tc}. Our main result is that all parity-even thermodynamic susceptibilities can be computed in terms of equilibrium two-point functions, while the earlier literature gave most of the susceptibilities in terms of equilibrium three-point functions. Explicitly, the Kubo formulas for parity-even susceptibilities are given by the equations (\ref{eq:f1-Kubo}) -- (\ref{eq:f7-Kubo}), and the Kubo formulas for parity-odd susceptibilities are given by (\ref{eq:f9-kubo}), (\ref{eq:f89-kubo}). The Kubo formulas are applicable both at zero and non-zero chemical potential.

The two-point functions are the zero-frequency equilibrium correlation functions in flat space, and can in principle be evaluated by the Euclidean methods, such as using lattice gauge theory. In fact, Ref.~\cite{Philipsen:2013nea} has already performed a lattice evaluation of $f_1$ (or rather $\kappa\equiv-2f_1$) in the $SU(3)$ Yang-Mills theory. We hope that the Kubo formulas derived in this paper will be useful for explicit calculations of the thermodynamic susceptibilities on the lattice as well as by holographic methods \cite{Baier:2007ix, Bhattacharyya:2008jc, Finazzo:2014cna} in strongly interacting quantum field theories.

The equilibrium constitutive relations for the energy-momentum tensor and the current are written down in Eqs.~(\ref{eq:Tmn}), (\ref{eq:cr-statics}), and (\ref{eq:Jmu}), (\ref{eq:NJ-cr-statics}), for a fluid not subject to external electromagnetic fields.%
\footnote{
   See \cite{Hernandez:2017mch} for fluids subject to a magnetic field (as would be relevant for magneto-hydrodynamics) and \cite{Kovtun:2016lfw} for fluids subject to both electric and magnetic fields (as would be relevant for polarized fluids).
}
The constitutive relations are written in the ``thermodynamic frame''~\cite{Jensen:2012jh}, which means that the fluid velocity in equilibrium is aligned with the timelike Killing vector, according to the definition~(\ref{eq:uTmu}). In principle, one can redefine the thermodynamic variables and write down the constitutive relations in the ``Landau-Lifshitz frame'', which corresponds to a redefinition of $T$, $\mu$, and $u^\alpha$, so that the expressions for ${\cal E}, {\cal N}, {\cal Q}^\mu$ written in terms of the new variables are made to look like ${\cal E} = \epsilon$, ${\cal N}=n$, ${\cal Q}^\mu = 0$. While doing so is fine at order $O(\partial^2)$, transforming to the Landau-Lifshitz frame (or any other frame) will also make the second-order susceptibilities $f_n(T,\mu)$ appear at  $O(\partial^3)$ and higher in the constitutive relations, confusing their true two-derivative nature.%
\footnote{
    Similarly, in 2+1 dimensions, writing the constitutive relations in the Landau-Lifshitz frame will make the $O(\partial)$ thermodynamic susceptibilities also appear at $O(\partial^2)$ and higher in the constitutive relations. The same comment applies to chiral anomalies: the thermodynamic frame is the frame in which the anomalous contributions to the constitutive relations only appear at one-derivative order~\cite{Banerjee:2012iz, Jensen:2013kka}. Writing down the constitutive relations in any other frame (such as the Landau-Lifshitz frame) will make anomalous terms appear at $O(\partial)$, $O(\partial^2)$, and higher in the constitutive relations, confusing their true one-derivative nature.
}
This makes any frame other than the thermodynamic frame ill-suited for a systematic understanding of thermodynamic contributions to the constitutive relations of $T^{\mu\nu}$ and $J^\mu$. Of course, the expectation values of $T^{\mu\nu}[g,A]$, $J^\mu[g,A]$ and the corresponding correlation functions are physical objects, and do not depend on one's choice of ``frame''.

\acknowledgments
We thank Guy Moore for helpful discussions. This work was supported in part by NSERC of Canada.

\appendix
\section{Translation of conventions}
\label{SS:TOC}
Several different notations and conventions for second-order transport coefficients exist in the literature. In this appendix we try to summarize some of the alternative conventions for second-order transport coefficients and their relation to the thermodynamic susceptibilities $f_n$ introduced in section~\ref{sec:thermodynamics}.

Early works on second-order hydrodynamics used the Landau-Lifshitz (LL) convention (also called ``frame''), which is a definition of the variables $T$, $\mu$, and $u^\alpha$ such that when $T^{\mu\nu}$ and $J^\mu$ are expressed in terms of the new variables using the decompositions (\ref{eq:Tmn}), (\ref{eq:Jmu}), one has ${\cal E}_L = \epsilon$, ${\cal N}_L=n$, and ${\cal Q}^\mu_L = 0$, where ``L'' signifies the LL frame. The transformations are $T_L = T+\delta T$, $\mu_L =\mu + \delta\mu$, $u_L^\alpha = u^\alpha + \delta u^\alpha$, where $\delta T$, $\delta\mu$, and $\delta u^\alpha$ contain terms $O(\partial)$ and higher. Explicitly, they are determined by (see e.g.~\cite{Kovtun:2012rj})
\begin{align*}
  & \epsilon_{,T}\delta T + \epsilon_{,\mu}\delta\mu  = {\cal E} - \epsilon\,,\\
  & n_{,T}\delta T + n_{,\mu}\delta\mu  = {\cal N} - n\,,\\
  & (\epsilon + p) \delta u^\mu = {\cal Q}^\mu \,,
\end{align*}
where $\epsilon(T,\mu)$ and $n(T,\mu)$ are defined below (\ref{eq:TJ0}), and ${\cal E}$, ${\cal N}$, ${\cal Q}^\mu$ are given by (\ref{eq:EE}), (\ref{eq:QQ}), (\ref{eq:NN}). Expressing (\ref{eq:Tmn}), (\ref{eq:Jmu}) in terms of $T_L$, $\mu_L$, and $u^\alpha_L$, one finds the equilibrium constitutive relations in the LL frame. The thermodynamic transport coefficients in the LL frame then emerge as combinations of the susceptibilities $f_n$ and their derivatives.

As an example, \cite{Bhattacharyya:2012nq} summarizes the constitutive relations for an uncharged fluid in the Landau-Lifshitz frame up to two-derivative terms. Comparing the constitutive relations in \cite{Bhattacharyya:2012nq} with the constitutive relations (\ref{eq:Tmn}) converted to the LL frame, we find 
\begin{subequations}
\label{eq:translation-1}
\begin{align}
  & \kappa_1 = - \, \frac{2}{T}\, f_1, \quad \kappa_2 = - \, \frac{2}{T}\, f_1' \,, \\
  &\lambda_3 = - \, \frac{2}{T}\, \big(f_1' - 4f_3\big), \quad \lambda_4 = \frac{1}{T} \, \big(4f_1'+2f_1''-2f_2\big) \,,\\
  &\zeta_2 = \frac{c_s^2}{T} \big( f_1 - f_1'\big) + \frac{1}{3T} \, f_1 \,,\\
  &\zeta_3 = -\,\frac{2 c_s^2}{T}  \,\big( f_2-f_1-f_1' \big) +\frac{2}{3T} \, \, \big( 2f_1'- f_1\big)\,,\\
  &\xi_3 = \frac{2 c_s^2}{T}  \,\big(f_1'-f_2-3f_3+f_3'\big) + \frac{2}{3T} \,\big(f_3 + 2f_1'\big)\,,\\
  &\xi_4 = -\,\frac{c_s^2}{T} \, \big( 4f_1' +2f_1''-f_2'-f_2\big) - \frac{1}{3T} \, \big( 4f_1''+2f_1'-f_2\big) \,.
\end{align}
\end{subequations}
The primes stand for $f_n' = T f_{n,T}$, $f_n'' = T^2 f_{n,T,T}$ (in an uncharged fluid), the comma denotes the derivative with respect to the argument that follows, and the speed of sound squared is $c_s^2 = \partial p/\partial\epsilon$.
The comparison with the LL-frame expressions was also performed in the original Refs.~\cite{Banerjee:2012iz, Jensen:2012jh}, using somewhat different conventions for the susceptibilities.

Ref.~\cite{Moore:2012tc} uses a different convention for the LL-frame transport coefficients for an uncharged fluid. Comparing the constitutive relations in \cite{Moore:2012tc} with the constitutive relations (\ref{eq:Tmn}) converted to the LL frame, we find
\begin{subequations}
\label{eq:translation-2}
\begin{align}
  & \kappa = -2 f_1, \quad \kappa^* = f_1' - 2f_1 \,,\\
  & \lambda_3 = 2(f_1' - 4f_3), \quad \lambda_4 = c_s^4\,\big( 4f_1' +2f_1''-2f_2 \big)\,,\\
  & \xi_3 = -2 c_s^2 \,\big( f_1'-f_2-3f_3+f_3' \big) - \coeff23\,\big( f_3 + 2f_1' \big)\,,\\
  & \xi_4 = - c_s^6 \, \big( 4f_1'+2f_1''-f_2'-f_2 \big) - \coeff13 c_s^4 \big( 4f_1'' + 2 f_1'-f_2 \big)\,,\\
  & \xi_5 = c_s^2\,\big( f_1 - f_1'\big) + \coeff13 f_1\,,\\
  & \xi_6 = -2 c_s^2\,\big( f_2-f_1-f_1'\big) + \coeff23 \, \big( 2f_1'-f_1 \big) \,.
\end{align}
\end{subequations}
These conversion formulas can be used to compare our results with those of \cite{Moore:2012tc}, which gives Kubo formulas for $\lambda_3$ and $\lambda_4$ in terms of three-point functions of $T^{\mu\nu}$.  

As an example, let us take $h_{tt}(x,y)$ as the only non-vanishing external source. Expanding the energy-momentum tensor (\ref{eq:Tmn}), (\ref{eq:cr-statics}) to $O(h^2)$ we find 
\begin{align}
\label{eq:Txyhtt}
  T^{xy} = 
  (f_1' - f_1)h_{tt,x,y}+
   \coeff12 \left[f_1'' + 2f_1' -2f_1 \right] h_{tt} h_{tt,x,y}
  +\coeff12\left[f_1'' + 3f_1' -f_1 - f_2 \right]h_{tt,x} h_{tt,y} \,.
\end{align}
Upon using the translation (\ref{eq:translation-2}), the first term in (\ref{eq:Txyhtt}) gives a Kubo formula for $\xi_5+\kappa/6$ (or $\kappa^* - \kappa/2$), while the last term in (\ref{eq:Txyhtt}) gives a Kubo formula for $\lambda_4/c_s^4 +2\kappa^* -\kappa$, in agreement with equation (A.39) in~\cite{Moore:2012tc}. As for the second term in (\ref{eq:Txyhtt}), it appears that equation (A.39) in \cite{Moore:2012tc} is missing~$f_1''$, though the term does not contain $\lambda_4$.

As another example, let us take $h_{ty}(z)$ as the only non-vanishing external source. Expanding the energy-momentum tensor (\ref{eq:Tmn}), (\ref{eq:cr-statics}) to $O(h^2)$ we find 
\begin{align}
\label{eq:Txxhty}
  T^{xx} = p + \left[f_3 - \coeff32 f_1 \right] h_{ty,z}^2 
  -2f_1 h_{ty} h_{ty,z,z} \,.
\end{align}
Upon using the translation (\ref{eq:translation-2}), one finds a Kubo formula for $\lambda_3$. 

Ref.~\cite{Megias:2014mba} presented the susceptibilities in a charged fluid, using the dimensionally reduced partition function, following the setup of~\cite{Banerjee:2012iz, Bhattacharyya:2014bha}. In order to compare our notation with that of~\cite{Megias:2014mba}, one can compare the partition functions directly, by applying the Kaluza-Klein reduction formulae of~\cite{Banerjee:2012iz} to our (\ref{eq:W2}), in the static gauge $V^\mu=(1,{\bf 0})$, $\Lambda_V=0$. One finds that the susceptibilities $M_i$ ($i=1,...,7$) and $N_k$ ($k=1,2$) of \cite{Megias:2014mba} are related to our $f_n$ by
\begin{align*}
  & M_1 = \frac{1}{T^3} \left( f_2 - 2f_1'\right) \,,\ \ \ \ 
    M_2 = T f_6\,,\ \ \ \ 
    M_3 = -\frac{1}{T}\left( f_7 + 2f_{1,\mu}\right) \,,\ \ \ \ 
    M_7 = \frac{f_1}{T}\,,\\
  & M_4 = \frac{1}{2T^3} \left( \coeff12 f_1 + \coeff12 f_3 + \mu^2 f_4 + \mu f_5 \right)\,,\ \ \ \ 
    M_5 = \frac{f_4}{2T}\,,\ \ \ \ 
    M_6 = \frac{1}{2T^2} \left( f_5 + 2\mu f_4 \right) \,,\\
  & N_1 = \frac{1}{2T}\left( \mu f_8 + 2f_9 \right)\,,\ \ \ \ 
    N_2 = \frac{f_8}{2T_0}\,,
\end{align*}
where as before $f_n'\equiv  T f_{n,T} + \mu f_{n,\mu}$, and the comma subscript denotes the partial derivative with respect to the argument that follows.
Note that the issue of ``frame'' transformations does not arise here, and the constitutive relations following from the partition function of \cite{Megias:2014mba} are in the same thermodynamic frame as ours.

\bibliographystyle{JHEP}
\bibliography{kubo-thermo}

\end{document}